\documentclass[aps,prx,amsmath,superscriptaddress,twocolumn,bibnotes,longbibliography,floatfix]{revtex4-2}
\usepackage{amsmath}
\usepackage{amsfonts}
\usepackage{graphicx}
\usepackage{color}
\usepackage{dsfont}
\usepackage{verbatim}
\usepackage{mathtools}
\usepackage[normalem]{ulem}
\usepackage{comment}
\usepackage{stackrel}
\usepackage{bm}
\usepackage{booktabs}

\usepackage[colorlinks=true, 
linkcolor=blue, 
urlcolor=blue,
citecolor=blue]{hyperref}
\usepackage{orcidlink}
\usepackage[linesnumbered]{algorithm2e}

\RestyleAlgo{ruled}
\SetKwComment{Comment}{/* }{ */}
\definecolor{myblue}{rgb}{.93, .93, 1}

\newcommand{\bsub}{\begin{subequations}}
	\newcommand{\esub}{\end{subequations}}

\newcommand{\vex}[1]{\bm{\mathrm{#1}}}

\begin{document}
	
	\title{Superconductivity from phonon-mediated retardation in a single-flavor metal}
	
	\author{Yang-Zhi~Chou~\orcidlink{0000-0001-7955-0918}}\email{yzchou@umd.edu}
	\affiliation{Condensed Matter Theory Center and Joint Quantum Institute, Department of Physics, University of Maryland, College Park, Maryland 20742, USA}
	
	\author{Jihang Zhu}
	\affiliation{Department of Materials Science and Engineering, University of Washington, Seattle, Washington 98195, USA}
	
	\author{Jay D. Sau}
	\affiliation{Condensed Matter Theory Center and Joint Quantum Institute, Department of Physics, University of Maryland, College Park, Maryland 20742, USA}

	\author{Sankar Das~Sarma}
	\affiliation{Condensed Matter Theory Center and Joint Quantum Institute, Department of Physics, University of Maryland, College Park, Maryland 20742, USA}
	
	\date{\today}
	
	\begin{abstract}
		We study phonon-mediated pairings in a single-flavor metal with a tunable Berry curvature. In the absence of Berry curvature, we discover an unexpected possibility: $p$-wave superconductivity emerging purely from the retardation effect, while the static BCS approximation fails to predict its existence. The gap function exhibits sign-change behavior in frequency (owing to the dynamical structure of the phonon-mediated interaction in the $p$-wave channel), and $T_c$ obeys a BCS-like scaling. We further show that the Berry curvature stabilizes the chiral $p$-wave superconductivity and can induce transitions to higher-angular-momentum pairings. Our results establish that the phonon-mediated mechanism is a viable pairing candidate in single-flavor systems, such as the quarter-metal superconductivity observed in rhombohedral graphene multilayers.
	\end{abstract}
	
	\maketitle
	
	\textit{Introduction. ---} Superconductivity (SC) is a cornerstone in modern condensed matter physics and arguably the most famous macroscopic quantum phenomenon. The defining properties of SC, vanishing electrical resistivity, expulsion of external magnetic field, and flux quantization, have been used in several practical applications. Most known superconducting materials can be explained by the formation of Cooper pairs via phonon-mediated pairing glue, as described by the Bardeen-Cooper-Schrieffer (BCS) theory \cite{BardeenJ1957}. Building on the BCS theory, the absence of vertex correction (i.e., Migdal theorem \cite{MigdalAB1958}) and full frequency dependence in the phonon-mediated interaction (i.e., Eliashberg theory \cite{EliashbergGM1960_ME}) are crucial for predicting $T_c$ in the realistic superconducting materials \cite{ChubukovAV2020a,MarsiglioF2020_review}. Due to the success of the BCS theory and Migdal-Eliashberg theory, the electron-phonon mechanism is often regarded as a well-understood and mundane mechanism for SC. Therefore, exploring unconventional, non-phonon superconducting materials is an active area of research, but any compelling evidence has been elusive \cite{DasSarmaS2025}.
	
	Recently, an intriguing quarter-metal (i.e., valley-polarized and spin-polarized) SC has been experimentally realized in rhombohedral graphene tetralayer and pentalayer \cite{HanT2025a}. Due to the valley polarization and the nontrivial Berry curvature in the normal state, $s$-wave pairing is forbidden, so it is natural to speculate that the pairing symmetry is chiral $p$-wave. It was pointed out by Refs.~\cite{ChouYZ2025,GeierM2025,YangH2025a} that the screened Coulomb interaction (i.e., Kohn-Luttinger mechanism \cite{KohnW1965}) may provide a potential explanation for the experiment. The phonon-mediated mechanism was not considered seriously because the static approximation (also called the BCS approximation) of the acoustic-phonon-mediated attraction is point-like \cite{WuF2019,ChouYZ2022b}, leaving no room for phonon-induced SC (s-wave or non-s-wave) in a single-flavor normal state (i.e., the quarter metal in rhombohedral graphene multilayers under a large displacement field) \cite{Phonon_pairing}. (The phonon mechanism is rarely considered as a mechanism for any non-s-wave SC in spite of the fact that almost all known SCs are phonon-mediated.) In fact, the actual phonon-mediated interaction is not strictly point-like, and the form factors of the interactions can induce nontrivial momentum dependence (e.g., in the presence of finite Berry curvature \cite{TanT2024a}), resulting in finite attractions in the triplet channels even within the static approximation \cite{WangYQ2024a,May-MannJ2025,PatriAS2025b,LiMR2025}. A conceptual question arises: Is the nontrivial interaction form factor or momentum dependence necessarily required for phonon-mediated odd-parity SC in a single-flavor model?
	
	In this Letter, we establish an unprecedented surprising possibility: Phonon-mediated odd-parity SC emerges in a single-flavor model purely driven by the dynamical structure of the phonon-mediated interaction, while the standard static BCS approximation predicts the absence of SC. The leading instability is of $p$-wave symmetry, with a gap function that shows sign-changing behavior and two characteristic dips at finite frequencies. The $T_c$, extracted by the linearized gap equation (LGE) with frequency dependence, exhibits a weak-coupling-BCS-like scaling $T_c\propto\exp(-1/\tilde{\lambda})$ with a modified dimensionless coupling constant $\tilde{\lambda}\approx\lambda/20.7$, where 
	$\lambda$ is the basic dimensionless coupling. In addition, we demonstrate that the presence of Berry curvature can enhance $T_c$ significantly and induce transitions to higher-odd-angular-momentum pairings. We discuss possible connections to the recent rhombohedral graphene multilayer experiments.

	\textit{Model. ---} We consider 2D single-flavor electrons (e.g., valley-polarized spin-polarized fermions) with a dispersion $E_{\vex{k}}$. The specific form of $E_{\vex{k}}$ is not essential here as long as it is rotationally invariant. For simplicity, we consider, without loss of generality, $E(\vex{k})=\vex{k}^2/(2m)$, where $m$ is the effective mass. In our model, electrons interact with the longitudinal acoustic phonons, whose dispersion is given by $\omega_{\vex{k}}=v_s|\vex{k}|$ with $v_s$ being the sound velocity. The effective inter-electron interactions are through emitting or absorbing phonons, resulting in an intrinsically dynamical interaction. The effective theory for electrons can be described by an imaginary-time action $\mathcal{S}=\mathcal{S}_0+\mathcal{S}_I$, where
	\begin{subequations}
		\begin{align}
			\label{Eq:S_0}\mathcal{S}_0=&\frac{1}{\beta}\sum_{\omega_n}\sum_{\vex{k}} \bar{c}_{\omega_n,\vex{k}}\left[-i\omega_n+E_{\vex{k}}-E_F\right]c_{\omega_n,\vex{k}},\\
			\label{Eq:S_I}\mathcal{S}_{I}=&\frac{-1}{2\beta \mathcal{A}}\sum_{\nu_n,\vex{q}}V(\nu_n,\vex{q})n(\nu_n,\vex{q})n(-\nu_n,-\vex{q}).
		\end{align}
	\end{subequations}
	In the above expressions, $c_{\omega_n,\vex{k}}$ is the fermionic field, $n(\nu_n,\vex{q})$ denotes the density, $\beta$ is the inverse temperature, $E_F$ is the Fermi energy, $\mathcal{A}$ is the 2D area, $V(\nu_n,\vex{q})=g\omega_{\vex{q}}^2/(\omega_{\vex{q}}^2+\nu_n^2)$ is the phonon-mediated interaction, and $g$ is the effective interaction strength. The density $n(\nu_n,\vex{q})$ is expressed by
	\begin{align}
		n(\nu_n,\vex{q}) =\frac{1}{\beta}\sum_{\vex{k},\vex{\omega_n}}\mathcal{F}_{\vex{k},\vex{k}+\vex{q}}\bar{c}_{\omega_n,\vex{k}}c_{\omega_n+\nu_n,\vex{k}+\vex{q}},
	\end{align}
	where $\mathcal{F}_{\vex{k},\vex{k}+\vex{q}}$ is a form factor due to projection onto the active band. In our theory, we consider an analytical form factor mimicking the ideal quantum geometry \cite{TanT2024a,May-MannJ2025,JahinA2026,BernevigBA2025,LiMR2025}
	\begin{align}\label{Eq:F_kk'}
		\mathcal{F}_{\vex{k},\vex{k}'}=e^{-\frac{\mathcal{B}}{4}|\vex{k}-\vex{k}'|^2-i\frac{\mathcal{B}}{2}\left(\vex{k}\times\vex{k}'\cdot\hat{z}\right)},
	\end{align}
	where $\mathcal{B}$ is the Berry curvature. This form factor is identical to that in the lowest Landau level, where Berry curvature is uniformly distributed. For our purpose, the uniformity is not essential, and we focus mainly on the Berry curvature near the Fermi surface. We treat $\mathcal{B}$ as a tunable parameter and choose $\mathcal{B}\ge 0$ without loss of generality. 
	
	\textit{Angular-momentum-dependent interaction and LGE. ---} To investigate SC, we focus on the BCS channel of the effective interaction described by
	\begin{align}
		\nonumber\mathcal{S}_I\rightarrow\frac{-1}{2\beta^3 \mathcal{A}}\sum_{\vex{k},\vex{k}'}\sum_{\omega_n,\omega_n'}&\mathcal{V}(\omega_n-\omega_n';\vex{k},\vex{k}')\\
		\label{Eq:S_I_BCS}&\times\bar{c}_{\omega_n,\vex{k}}\bar{c}_{-\omega_n,-\vex{k}}c_{-\omega_n',-\vex{k}'}c_{\omega_n',\vex{k}'},
	\end{align}
	where $\mathcal{V}(\nu_n;\vex{k},\vex{k}')\equiv V(\nu_n,\vex{k}-\vex{k}')\mathcal{F}_{\vex{k},\vex{k}'}\mathcal{F}_{-\vex{k},-\vex{k}'}$ is the pairing interaction potential incorporating the form factor contribution. 
	
	\begin{figure}[t]
		\includegraphics[width=0.45\textwidth]{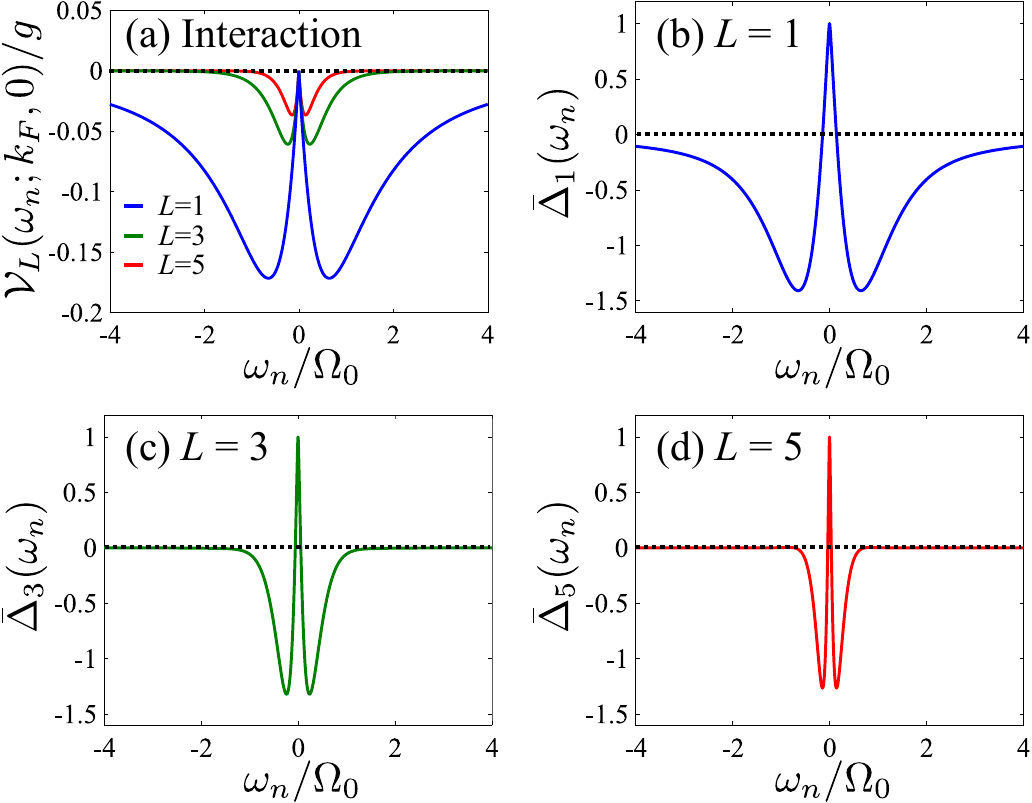}
		\caption{Angular momentum $L$-channel interactions and gap functions for $\mathcal{B}=0$. (a) The dimensionless interaction $\mathcal{V}_L(\nu_n;k_F,0)/g$ of $L=1,3,5$. The negative value means repulsion due to the convention for pairing interaction. (b)-(d) The rescaled gap function $\bar{\Delta}_L(\omega_n)\equiv\Delta_L(\omega_n)/\Delta_L(\pi T)$ as a function of $\omega_n$. We keep $T_c=10^{-3}\Omega_0$ and use different values of $\lambda$: (b) $\lambda=3.2431$ for $L=1$; (c) $\lambda=11.0439$ for $L=3$; (d) $\lambda=20.3124$ for $L=5$. The black dashed lines indicate zero on the $y$-axis. $4000$ Matsubara frequencies are included in the calculations.
		}
		\label{Fig:Int_and_gap}
	\end{figure}
	
	Assuming that the dominant scattering events happen near the Fermi level, we then project the interaction onto the circular Fermi surface. Thus, we can approximate
	\begin{align}
		V(\nu_n,\vex{k}-\vex{k}')\approx&g\frac{1-\cos\left(\theta_{\vex{k}}-\theta_{\vex{k}'}\right)}{1-\cos\left(\theta_{\vex{k}}-\theta_{\vex{k}'}\right)+\nu_n^2/\Omega_0^2},\\
		\mathcal{F}_{\vex{k},\vex{k}'}\mathcal{F}_{-\vex{k},-\vex{k}'}\approx&\exp\left[-\mathcal{B}k_F^2+\mathcal{B}k_F^2e^{i\left(\theta_{\vex{k}}-\theta_{\vex{k}'}\right)}\right],
	\end{align}
	where $k_F$ is the Fermi wavevector, $\Omega_0=\sqrt{2}v_sk_F$ is the characteristic phonon frequency, and $\theta_{\vex{k}}$ is the angle between $\vex{k}$ and the x axis. Finally, we derive the approximate phonon-mediated interaction potential and the corresponding angular-momentum expansion
	\begin{align}
		\mathcal{V}(\nu_n;\vex{k},\vex{k}')
		\approx\sum_Le^{iL(\theta_{\vex{k}}-\theta_{\vex{k}'})}\mathcal{V}_L(\nu_n;k_F,\mathcal{B}),
	\end{align}
	where $L$ denotes the angular momentum,
	\begin{align}
		\label{Eq:V_L}\mathcal{V}_L(\nu_n;k_F,\mathcal{B})\!=&g\!\left[b_L(\mathcal{B}k_F^2)\!-\!\sum_{m=0}^{\infty}b_m(\mathcal{B}k_F^2)R_{L-m}(\nu_n/\Omega_0)\right],\\
		\label{Eq:b_m}b_m(x\ge 0)=&\begin{cases}
			\frac{x^m}{m!}e^{-x} & \text{ for }m> 0,\\[1mm]
			e^{-x} & \text{ for }m= 0,\\[1mm]
			0 & \text{ for }m<0,
		\end{cases}\\
		R_m(\varpi)=&\frac{\varpi^2\!\left[\left(1\!+\!\varpi^2\right)\!-\!\sqrt{\left(1\!+\!\varpi^2\right)^2\!-\!1}\right]^{|m|}}{\sqrt{\left(1+\varpi^2\right)^2-1}}.
	\end{align}
	In the above expressions, $b_m(\beta k_F^2)$ is obtained from Fourier decomposition of the $\mathcal{F}_{\vex{k},\vex{k}'}\mathcal{F}_{-\vex{k},-\vex{k}'}$ term, and $R_L(\nu_n/\Omega_0)$ is associated with the $V(\nu_n,\vex{k}-\vex{k}')$. The standard static BCS approximation for the phonon-mediated interaction corresponds to setting $\nu_n=0$. Since $R_m(\varpi)$ vanishes linearly as $\varpi\rightarrow 0$, the static BCS approximation yields $\mathcal{V}_L(0;k_F,\mathcal{B})=gb_L(\mathcal{B}k_F^2)\ge 0$, where Berry curvature crucially determines the static attraction. The essence of this work is to treat the full frequency dependence in the pairing interactions. 
	
	To incorporate the dynamical structure of the phonon-mediated interactions, we derive the frequency-dependent LGE for the angular-momentum-$L$ pairing (see SM for a derivation \cite{SM}), which is equivalent to the Eliashberg theory \cite{EliashbergGM1960_ME,ChubukovAV2020a,MarsiglioF2020_review} without the quasiparticle renormalization. Assuming $E_F\gg\Omega_0$ (i.e., Migdal theorem applies), the LGE is given by
	\begin{align}\label{Eq:LGE}
		\Delta_L(\omega_n)=\frac{\rho_0\pi}{\beta}\sum_{\omega_n'}\frac{\mathcal{V}_L(\omega_n-\omega_n';k_F,\mathcal{B})}{|\omega_n'|}\Delta_L(\omega_n'),
	\end{align}
	where $\rho_0$ is the density of state at the Fermi level and $\Delta_L$ is the gap function associated with angular momentum $L$, expressed by
	\begin{align}
		\nonumber\Delta_L(\omega_n)=\frac{1}{\beta^2 \mathcal{A}}\sum_{\omega_n'}\sum_{\vex{k}'}&\,\mathcal{V}_L(\omega_n\!-\!\omega_n';k_F,\mathcal{B})\\
		\label{Eq:Delta_L}&\times e^{-iL\theta_{\vex{k}'}}\!\left\langle c_{-\omega_n',-\vex{k}'}c_{\omega_n',\vex{k}'}\right\rangle.
	\end{align}
	Equation (\ref{Eq:LGE}) can be viewed as an eigenvalue problem with frequencies being the matrix indices. The solution corresponds to the right eigenvalue equaling to unity. We first focus on $\mathcal{B}=0$ and then discuss situations with $\mathcal{B}>0$.

	\begin{figure}[t]
		\includegraphics[width=0.3\textwidth]{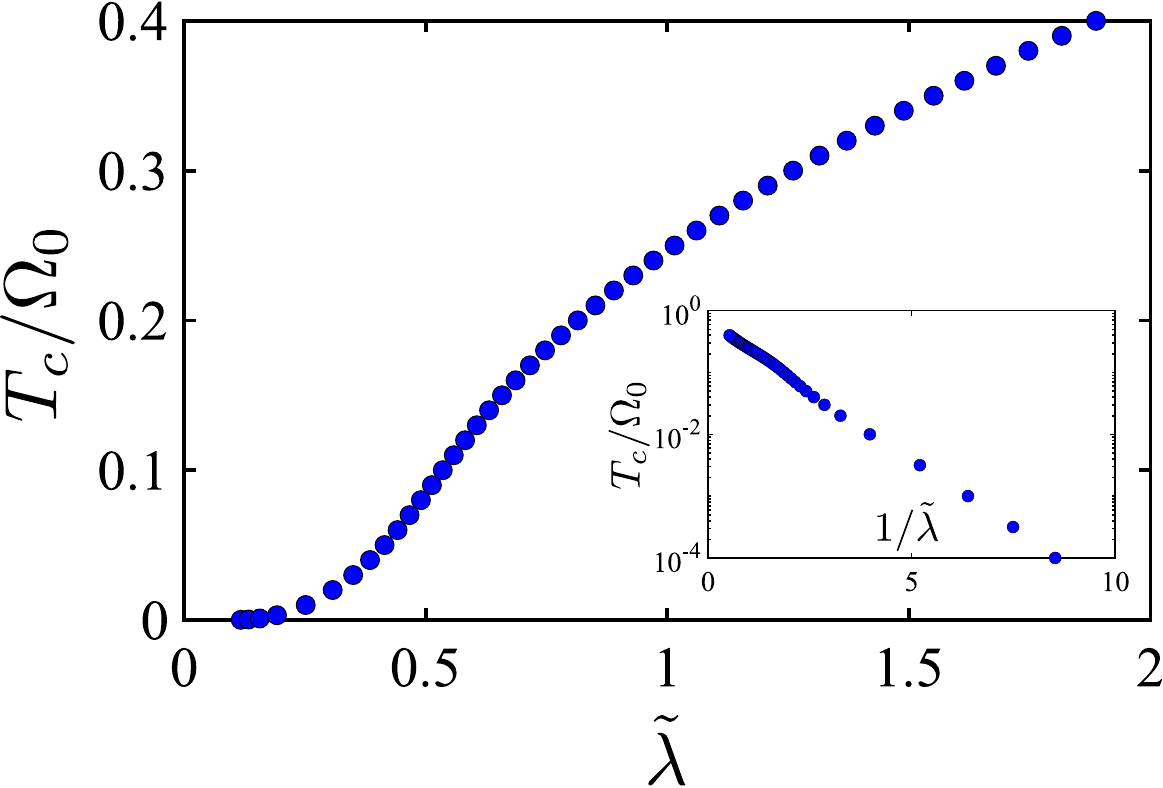}
		\caption{$T_c$ of $p$-wave SC as a function of the rescaled coupling constant $\tilde{\lambda}$ for $\mathcal{B}=0$. The dimensionless coupling constant $\tilde{\lambda}$ is defined by $\tilde{\lambda}\equiv\lambda/20.7$. Inset: $T_c$ in the logarithmic scale as a function of $1/\tilde{\lambda}$, showing the BCS-like scaling, $T_c\propto\exp(-1/\tilde{\lambda})$.
		}
		\label{Fig:Tc_L1}
	\end{figure}
	
	\textit{Phonon-mediated SC with $\mathcal{B}=0$. ---} In the absence of Berry curvature ($\mathcal{B}=0$), the interaction of the angular-momentum-$L$ channel becomes
	\begin{align}
		\label{Eq:V_L_no_B}\mathcal{V}_L(\nu_n;k_F,0)=g\left[\delta_{L,0}-R_{L}(\nu_n/\Omega_0)\right].
	\end{align} 
	For $L=0$, $\mathcal{V}_0(\nu_n;k_F,0)>0$ for all $\nu_n$, suggesting attraction. However, the even-angular-momentum pairing is not allowed in our model due to the Pauli exclusion principle. One can straightforwardly show that $\mathcal{V}_{L\neq 0}(\nu_n;k_F,0)\le 0$ for all $\nu_n$, i.e., no attraction between electrons at all frequencies. In Fig.~\ref{Fig:Int_and_gap}(a), we plot $\mathcal{V}_{L}$ with $\mathcal{B}=0$ for $L=1,3,5$, showing the non-monotonic dependence in frequency and $\mathcal{V}_{L}(0;k_F,0)= 0$. Thus, the static BCS approximation predicts the absence of SC (without Berry curvature). However, the dynamical structure of the $\mathcal{V}_L(\nu_n;k_F,0)$ is nontrivial and should be investigated carefully. A well-established example is the seminal work by Morel and Anderson \cite{MorelP1962}, which demonstrates the existence of SC despite a repulsive pairing interaction at all frequencies. Retardation by itself can produce pairings even if the static BCS limit predicts no SC!

	With Eqs.~(\ref{Eq:V_L_no_B}) and (\ref{Eq:LGE}), we derive the LGE for $\mathcal{B}=0$, given by
	\begin{align}\label{Eq:LGE_no_B}
		\Delta_L(\omega_n)=\frac{\lambda\pi}{\beta}\sum_{\omega_n'}\frac{\left[\delta_{L,0}-R_L((\omega_n-\omega_n')/\Omega_0)\right]}{|\omega_n'|}\Delta_L(\omega_n'),
	\end{align}
	where $\lambda=g\rho_0$.
	We focus only on the odd-$L$ even-frequency pairings in this Letter. The odd-frequency pairings are known to be sensitive to quasiparticle renormalization \cite{LinderJ2019,PimenovD2022,LangmannE2022}, a subject beyond the scope of our current study. We analyze the odd-$L$ even-frequency pairings by solving Eq.~(\ref{Eq:LGE_no_B}) with a frequency cutoff $\Lambda\gg\Omega_0$. The results converge well for $\Lambda>5\Omega_0$ because $\mathcal{V}_{L}(\nu_n;k_F,0)$ decays to zero rapidly for $|\nu_n|>5\Omega_0$. See SM \cite{SM} for a discussion on convergence.
	
	We find that the leading SC instability is $p$-wave, and $L=1$ and $L=-1$ are degenerate. As plotted in Fig.~\ref{Fig:Int_and_gap}(b), the gap function for $L=1$ exhibits a peak at zero frequency, sign-changing behavior, two dips (the most negative locations) track the two maxima of $|\mathcal{V}_1(\nu_n;k_F,0)|$ [see the blue curve in Fig.~\ref{Fig:Int_and_gap}(a)]. The sign-changing behavior is expected for a non-attractive pairing potential because a minus sign is needed for compensating the minus sign in front of $R_L$ in Eq.~(\ref{Eq:LGE_no_B}). This sign-reversal feature is reminiscent of the Bogoliubov-Tomachov-Morel-Anderson pairing potential \cite{BogoljubovNN1958,MorelP1962} (i.e., a dynamical phonon-mediated attraction with a large static Coulomb repulsion), but the Coulomb interaction is absent in our case. Moreover, the existence of SC here cannot be understood just by the Morel-Anderson renormalization (i.e., the $\mu^*$ effect, marginally irrelevant renormalization group flows for repulsive interactions in the BCS channel) \cite{MorelP1962,ColemanP2015_book}. The two dips at finite frequencies are also nontrivial as they inherit the dynamical structure of $\mathcal{V}_1(\nu_n;k_F,0)$. As shown in Figs.~\ref{Fig:Int_and_gap}(c)-(d), similar characteristics manifest in the gap functions of $L=3$ and $L=5$, so they are also allowed in principle.

	Besides the gap function, we also investigate the $T_c$ as a function of $\lambda$. We solve the LGE keeping the frequency cutoff $\Lambda>5\Omega_0$ for $10^{-4}\Omega_0\le T_c\le 0.4\Omega_0$. As shown in Fig.~\ref{Fig:Tc_L1}, we obtain the weak-coupling-BCS-like scaling: $T_c\propto \exp(-1/\tilde{\lambda})$ for the leading $L=1$ pairing, where numerically $\tilde{\lambda}=\lambda/20.7$ is a modified dimensionless coupling constant. The manifestation of the BCS-like scaling suggests that SC emerges for an arbitrarily small $\lambda$ at zero temperature, but the $T_c$ is generically much smaller than the conventional BCS spin-singlet $s$-wave SC (because $\tilde{\lambda}\ll\lambda$). We expect that the scaling $T_c\propto \exp(-1/\tilde{\lambda})$ should survive the full Eliashberg treatment (incorporating quasiparticle renormalization) for $\lambda<1$. For $\lambda>1$, the quasiparticle renormalization will likely suppress the $T_c$ and result in a different dependence on $\tilde{\lambda}$, as is well-known in the theories of strong-coupling SC \cite{ChubukovAV2020a,MarsiglioF2020_review}.
	
	\begin{figure}[t!]
	\includegraphics[width=0.45\textwidth]{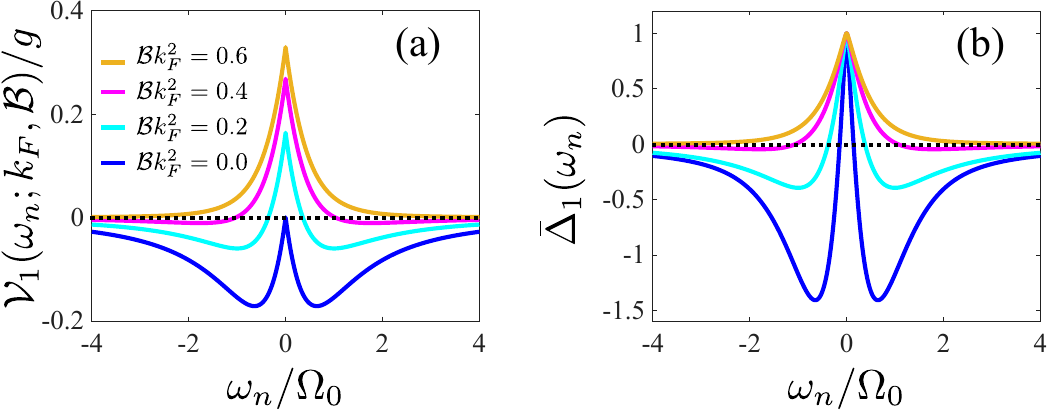}
	\caption{$L=1$ interaction potential and gap functions with small $\mathcal{B}k_F^2$. (a) The $L=1$ pairing dimensionless interaction $\mathcal{V}_1/g$ as a function of frequency with several representative values of $\mathcal{B}k_F^2$. (b) The corresponding rescaled gap function $\bar{\Delta}_L(\omega_n)\equiv\Delta_L(\omega_n)/\Delta_L(\pi T)$ as a function of $\omega_n$. We keep $T_c=10^{-3}\Omega_0$ and use different values of $\mathcal{B}k_F^2$. $\lambda=3.2431$ for $\mathcal{B}k_F^2=0$ (blue curve); $\lambda=1.3468$ for $\mathcal{B}k_F^2=0.2$ (cyan curve); $\lambda=0.7591$ for $\mathcal{B}k_F^2=0.4$ (magenta curve); $\lambda=0.5888$ for $\mathcal{B}k_F^2=0.6$ (gold curve).
	}
	\label{Fig:Gap_fcn_with_B}
\end{figure}		
	
	\textit{Phonon-mediated SC with $\mathcal{B}>0$. ---} Now, we discuss the influence of a nontrivial Berry curvature, specifically $\mathcal{B}>0$. The presence of $\mathcal{B}$ breaks the degeneracy between $L$ and $-L$ channels, and the $L>0$ pairings are generically favored. (See SM for a discussion \cite{SM}.) Thus, we focus on $L=1,3,5$, corresponding to the three leading odd-parity even-frequency pairings.

	With $\mathcal{B}> 0$, the static part of the pairing interaction, $\mathcal{V}_L(0;k_F,\mathcal{B})=gb_L(\mathcal{B}k_F^2)$, becomes positive, suggesting a static attraction and existence of SC. This is a very different situation from the $\mathcal{B}=0$ case, where pairing interactions are always non-attractive. In Fig.~\ref{Fig:Gap_fcn_with_B}, we plot $\mathcal{V}_1(\omega_n;k_F,\mathcal{B})$ and the $L=1$ gap functions with small $\mathcal{B}k_F^2$. The pairing potential $\mathcal{V}_1(\omega_n;k_F,\mathcal{B})$ acquires a noticeable static attraction even with $\mathcal{B}k_F^2=0.2$, and the corresponding gap function shows a wider peak width at zero frequency, reduced dips (the most negative positions at finite frequencies), and the movement of zeros toward $\pm \infty$ (roughly tracking the zeros in the pairing interaction). As $\mathcal{B}k_F^2$ increases to $0.6$, the pairing potential $\mathcal{V}_1(\omega_n;k_F,\mathcal{B})$ becomes positive (i.e., attractive) at all frequencies in our calculations. Concomitantly, the dips and zeros disappear completely, suggesting the absence of the sign-changing feature. The $\mathcal{B}k_F^2=0.6$ case is similar to the conventional phonon-mediated $s$-wave SC without Coulomb suppression. In this scenario, the static BCS approximation of phonon-mediated pairing should qualitatively capture the properties of SC. The results in Fig.~\ref{Fig:Gap_fcn_with_B} also imply that adding Berry curvature can substantially enhance the phonon-mediated SC since the required $\lambda$ for $T_c=10^{-3}\Omega_0$ is smaller for a larger Berry curvature.

\begin{figure}[t]
	\includegraphics[width=0.3\textwidth]{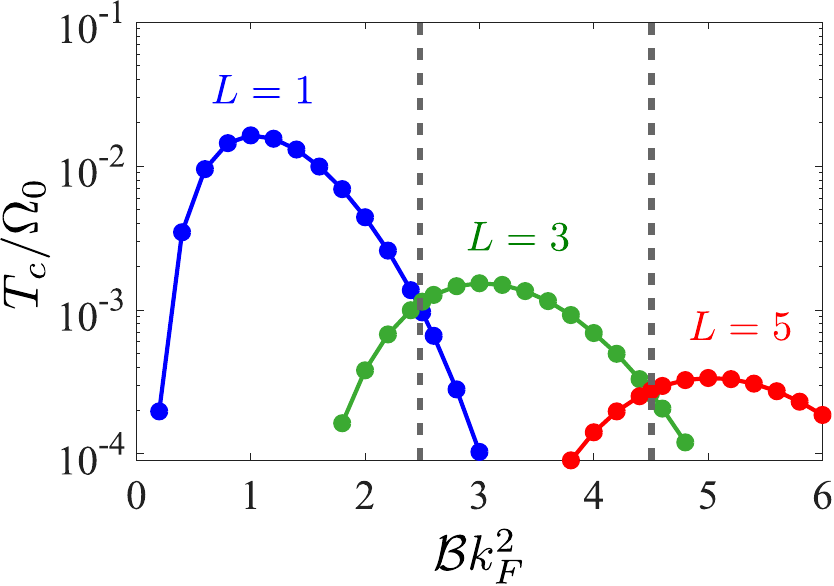}
	\caption{$T_c$ as a function of $\mathcal{B}k_F^2$ for $L=1,3,5$ (with $\lambda=1$). Note that the $T_c/\Omega_0$ is in the logarithmic scale. The results show that $T_c$'s are nonmonotonic as tuning $\mathcal{B}k_F^2$. See main text for a detailed discussion. Blue dots: $L=1$; green dots: $L=3$; red dots: $L=5$. The dashed lines indicate the phase transitions between different angular momentum $L$.
	}
	\label{Fig:Tc_vs_B}
\end{figure}

	Notably, the effect due to Berry curvature is non-monotonic because $\mathcal{V}_L(0;k_F,\mathcal{B})=gb_L(\mathcal{B}k_F^2)$ is maximized at $\mathcal{B}k_F^2=L$ for $L>0$. To investigate how SC is influenced by the finite Berry curvature, we solve LGE and extract $T_c$ by varying $\mathcal{B}k_F^2$ for $L=1,3,5$ and plot the results (setting $\lambda=1$) in Fig.~\ref{Fig:Tc_vs_B}. For the $L=1$ case, small $\mathcal{B}$ significantly boosts $T_c$ and the maximal $T_c$ is achieved at $\mathcal{B}k_F^2=1$. Similar non-monotonic behaviors are also observed in the cases $ L=3$ and $ L=5$. With the results mentioned above, a phase diagram is shown in Fig.~\ref{Fig:Tc_vs_B}. We find the dominant $L=1$ pairing for $0\le\mathcal{B}k_F^2<2.5$, $L=3$ pairing for $2.5<\mathcal{B}k_F^2<4.5$, and $L=5$ for $4.5<\mathcal{B}k_F^2<6$. (We do not investigate $\mathcal{B}k_F^2>6$ or $L>5$ because the $T_c$ is numerically inaccessible.) The phase boundaries are roughly at $\mathcal{B}k_F^2=2.5$ and $\mathcal{B}k_F^2=4.5$ (gray dashed lines in Fig.~\ref{Fig:Tc_vs_B}), close to the estimate based on $\mathcal{V}_L(0;k_F,\mathcal{B})=gb_L(\mathcal{B}k_F^2)$: $\mathcal{B}k_F^2=\sqrt{6}$ and $\mathcal{B}k_F^2=\sqrt{20}$. Since $\mathcal{V}_L(0;k_F,\mathcal{B})\propto\exp(-\mathcal{B}k_F^2)$ (up to power-law corrections), $T_c$ is suppressed for a large $\mathcal{B}k_F^2$, showing the detrimental effect of large Berry curvature. The results here are consistent with Ref.~\cite{May-MannJ2025}, but we incorporate the full dynamical structure of phonon-mediated attraction and extract $T_c$ explicitly. Similar results of $T_c$ with finite Berry curvature (e.g., $T_c$ enhanced by small Berry curvature and the nonmonotonic behavior of $T_c$) are also found in the Kohn-Luttinger induced SC \cite{JahinA2026,May-MannJ2025,ShavitG2025}.
	
	The results presented here are based on the ideal-quantum-geometry form factor $\mathcal{F}_{\vex{k},\vex{k}'}$ \cite{TanT2024a,May-MannJ2025,BernevigBA2025,LiMR2025} given by Eq.~(\ref{Eq:F_kk'}). Away from the ideal quantum geometry, we still expect that the $L>0$ pairings are favored (assuming positive Berry curvature), but the Berry curvature distribution can become nonuniform. The qualitative results in Fig.~\ref{Fig:Tc_vs_B} should hold in general (with the ``tuning parameter'' replaced by the averaged $\mathcal{B}k_F^2$ over the Fermi surface).
	
	\textit{Discussion. ---} The model considered in this Letter is idealized and ignores several material-specific details \cite{VP}. The major assumption in the frequency-dependent LGE [Eq.~(\ref{Eq:LGE})] is the negligible quasiparticle renormalization. Solving the full Migdal-Eliashberg equations with the pairing interaction $\mathcal{V}_L$ [Eq.~(\ref{Eq:V_L})] is likely nontrivial as the gap function manifests sign-changing and other features. These dynamical structures of the gap function can be in principle measured through the tunneling spectroscopy. We expect quasiparticle renormalization to have quantitative effects, but our qualitative result on the importance of retardation would still apply.
	Another assumption is the projection of the pairing interaction onto the circular Fermi surface, which is valid for a frequency cutoff $\Lambda\ll E_F$ and a rotationally symmetric $E_{\vex{k}}$. For materials with complex band structures, solving the Eliashberg equation with full momentum and frequency dependence may be necessary. The main goal of this Letter is to uncover the existence of the single-flavor SC purely driven by the dynamical phonon-mediated interaction, which cannot be understood by the static BCS approximation or by the Morel-Anderson renormalization (i.e., the $\mu^*$ effect) \cite{MorelP1962}.

	We discuss several implications of our theory for the recent rhombohedral graphene multilayer experiments \cite{HanT2025a,MorissetteE2025b}, which show quarter-metal SC \cite{Stripe_SC}. Focusing on the tetralayer graphene, we estimate the dimensionless coupling constant $\lambda\approx g\rho_0\approx3.13$, using $g=0.474$ eV nm$^2$ \cite{WuF2019,ChouYZ2021} and $\rho_0=6.6$ eV$^{-1}$ nm$^{-2}$ (corresponding to $n_e\approx 0.52\times 10^{12}$ cm$^{-2}$ inside the experimental SC1 region \cite{HanT2025a} with a single Fermi surface \cite{SM}). As shown in SM \cite{SM}, we compute $\mathcal{B}k^2$ for the $k$ points near the single Fermi surface using the $k\cdot p$ model \cite{GhazaryanA2023} and find the averaged value is $0.47$, implying an $L=1$ pairing interaction based on Fig.~\ref{Fig:Tc_vs_B}. Similar averaged $\mathcal{B}k_F^2$ values are obtained for the pentalayer ($1.14$) and hexalayer ($1.5$), suggesting dominant $L=1$ pairings, too. As shown in Fig.~\ref{Fig:Tc_vs_B}, the $T_c$ of $L=1$ pairing is significantly boosted for $0.4<\mathcal{B}k_F^2<2$. With the estimated value of $\lambda$ and the averaged values of $\mathcal{B}k_F^2$, we conclude that phonon-mediated pairing glues are significant for SC in the rhombohedral graphene multilayer experiments. One potential issue is that our theory is based on projection onto the Fermi surface and assuming the Migdal theorem, which are not strictly applicable to the rhombohedral graphene systems. 
	
	Another interesting numerical observation is that the value of the averaged $\mathcal{B}k_F^2$ monotonically increases as the layer number ($n$) increases. Therefore, the $L>1$ pairings may be realized in rhombohedral multilayers with layer number $n>6$. Meanwhile, large $\mathcal{B}k_F^2$ also suppresses SC because the static pairing attraction scales as $\exp(-\mathcal{B}k_F^2)$ (up to power-law corrections).
	Since the averaged $\mathcal{B}k_F^2$ likely increases further for larger layer numbers ($n>6$), we speculate that the observable quarter-metal SC only exists in a finite range of $n\ge 4$, not all the way to the rhombohedral graphite limit (i.e., $n\rightarrow\infty$).
	
	Finally, we comment on the interplay between the Kohn-Luttinger (screened Coulomb interaction) and phonon-mediated mechanisms in the quarter-metal chiral SC \cite{HanT2025a}. As discussed in a number of studies \cite{GeierM2025,YangH2025a,May-MannJ2025,QinQ2024,TavakolO2025}, the pairings due to Coulomb interaction favor $L=-1$ pairing with $\mathcal{B}>0$, the opposite pairing chirality predicted by the phonon-mediated mechanism. Note that the two mechanisms actually cooperate rather than compete, as they contribute to both $L=1$ and $L=-1$ channels, and the dominant mechanism determines the precise chirality. Regardless of the dominant pairing mechanism, our theory establishes that phonon-mediated pairings could be relevant in the unconventional SC of rhombohedral graphene multilayer systems.
	
	\begin{acknowledgments}
		We thank Andrei Bernevig for useful discussion.
		This work is supported by the Laboratory for Physical Sciences (Y.-Z.C., J.D.S, and S.D.S), by the Joint Quantum Institute (J.D.S.), and by the U.S. Department of Energy, Office of Basic Energy Sciences, under Contract No. DE-SC0025327. (J.Z.).
		
		\textit{Data availability--} The data used in the figures of this work is openly available at \cite{Chou2026Data}.
	\end{acknowledgments}

	%\bibliography{Phonon_SC}

\begin{thebibliography}{38}
\bibitem [{VP()}]{VP}%
		\BibitemOpen
		\href@noop {} {}\bibinfo {note} {For example, the valley-polarized metal in a
			graphene multilayer intrinsically carries a finite momentum, so is the
			resulting superconductivity. Thus, the consequences of a finite-momentum
			pairing cannot be captured by our model}\BibitemShut {NoStop}%
\end{thebibliography}
	
	%apsrev4-2.bst 2019-01-14 (MD) hand-edited version of apsrev4-1.bst
	%Control: key (0)
	%Control: author (8) initials jnrlst
	%Control: editor formatted (1) identically to author
	%Control: production of article title (0) allowed
	%Control: page (0) single
	%Control: year (1) truncated
	%Control: production of eprint (0) enabled
	%

	\newpage \clearpage 
	
	\onecolumngrid
	
	\begin{center}
		{\large
			Superconductivity from phonon-mediated retardation in a single-flavor metal
			\vspace{4pt}
			\\
			SUPPLEMENTAL MATERIAL
		}
	\end{center}
	
	\setcounter{figure}{0}
	\renewcommand{\thefigure}{S\arabic{figure}}
	\setcounter{equation}{0}
	\renewcommand{\theequation}{S\arabic{equation}}

	In this supplemental material, we provide technical details for the main results presented in the main text.
	
	\section{Angular momentum decomposition of phonon-mediated interaction}
	
	In this section, we provide derivations of $\mathcal{V}_L(\nu_n;k_F,\mathcal{B})$, the pairing interaction of angular-momentum $L$. As discussed in the main text, the interaction in the BCS channel (after incorporating the form factor) is
	\begin{align}
		\mathcal{V}(\nu_n;\vex{k},\vex{k}')\equiv V(\nu_n,\vex{k}-\vex{k}')\mathcal{F}_{\vex{k},\vex{k}'}\mathcal{F}_{-\vex{k},-\vex{k}'}
		\approx\sum_Le^{iL(\theta_{\vex{k}}-\theta_{\vex{k}'})}\mathcal{V}_L(\nu_n;k_F,\mathcal{B}),
	\end{align}
	where $\mathcal{V}_L(\nu_n;k_F,\mathcal{B})$ is the Fourier component of angular momentum $L$. First, we derive the projected interaction onto the circular Fermi surface. Then, we derive the corresponding angular decomposition.
	
	\subsection{Projection onto circular Fermi surface}
	
	We assume that the dominant scattering events occur near the Fermi level, which is valid when $E_F$ is the dominant scale. We then treat $\vex{k}$ and $\vex{k}'$ as two vectors on a circle with radius $k_F$, whose angles are $\theta_{\vex{k}}$ and $\theta_{\vex{k}'}$, respectively. Then, we derive the following identities
	\begin{align}
		|\vex{k}-\vex{k}'|^2=&2k_F^2-2k_F^2\cos\left(\theta_{\vex{k}}-\theta_{\vex{k}'}\right),\\
		\vex{k}\times\vex{k}' \cdot\hat{z}=&-k_F^2\sin\left(\theta_{\vex{k}}-\theta_{\vex{k}'}\right).
	\end{align}
	With the above identities, we derive the projected bare phonon-mediated interaction and form factors as follows:
	\begin{align}
		V(\nu_n,\vex{k}-\vex{k}')=&g\frac{v_s^2|\vex{k}-\vex{k}'|^2}{v_s^2|\vex{k}-\vex{k}'|^2+\nu_n^2}\approx g\frac{2v_s^2k_F^2\left[1-\cos\left(\theta_{\vex{k}}-\theta_{\vex{k}'}\right)\right]}{2v_s^2k_F^2\left[1-\cos\left(\theta_{\vex{k}}-\theta_{\vex{k}'}\right)\right]+\nu_n^2}=g\frac{1-\cos\left(\theta_{\vex{k}}-\theta_{\vex{k}'}\right)}{1-\cos\left(\theta_{\vex{k}}-\theta_{\vex{k}'}\right)+\nu_n^2/\Omega_0^2},\\
		\nonumber\mathcal{F}_{\vex{k},\vex{k}'}\mathcal{F}_{-\vex{k},-\vex{k}'}=&e^{-\frac{\mathcal{B}}{4}|\vex{k}-\vex{k}'|^2-i\frac{\mathcal{B}}{2}\left(\vex{k}\times\vex{k}'\cdot\hat{z}\right)}e^{-\frac{\mathcal{B}}{4}|(-\vex{k})-(-\vex{k}')|^2-i\frac{\mathcal{B}}{2}\left[(-\vex{k})\times(-\vex{k}')\cdot\hat{z}\right]}\approx e^{-\mathcal{B}k_F^2\left[1-\cos\left(\theta_{\vex{k}}-\theta_{\vex{k}'}\right)\right]+i\mathcal{B}k_F^2\sin\left(\theta_{\vex{k}}-\theta_{\vex{k}'}\right)}\\
		=&\exp\left[-\mathcal{B}k_F^2+\mathcal{B}k_F^2e^{i\left(\theta_{\vex{k}}-\theta_{\vex{k}'}\right)}\right],
	\end{align}
	where $\Omega_0=\sqrt{2}v_sk_F$ is the characteristic phonon frequency.
	The above expressions are the approximate forms used in the main text.

	\subsection{Angular decomposition}
	
	Now, we derive the angular decomposition of the effective pairing interaction. The phonon-mediated kernel can be expressed by
	\begin{align}
		V(\nu_n,\vex{k}-\vex{k}')\approx& g\frac{1-\cos\left(\theta-\theta'\right)}{1-\cos\left(\theta-\theta'\right)+\nu_n^2/\Omega_0^2}
		=g-g\frac{\nu_n^2}{\Omega_0^2}\frac{1}{\left(1+\nu_n^2/\Omega_0^2\right)-\cos\left(\theta-\theta'\right)}\\
		=&g-g\frac{\nu_n^2}{\Omega_0^2}\sum_{m=-\infty}^{\infty}e^{im(\theta-\theta')}\frac{\left[\left(1+\nu_n^2/\Omega_0^2\right)-\sqrt{\left(1+\nu_n^2/\Omega_0^2\right)^2-1}\right]^{|m|}}{\sqrt{\left(1+\nu_n^2/\Omega_0^2\right)^2-1}}\\
		\equiv&g\sum_{m=-\infty}^{\infty}e^{im(\theta-\theta')}\left[\delta_{m,0}-R_m(\nu_n/\Omega_0)\right]
	\end{align}
	where
	\begin{align}
		R_m(\varpi)=&\frac{\varpi^2\!\left[\left(1\!+\!\varpi^2\right)\!-\!\sqrt{\left(1\!+\!\varpi^2\right)^2\!-\!1}\right]^{|m|}}{\sqrt{\left(1+\varpi^2\right)^2-1}}
	\end{align}
	
	To derived the Fourier transform, we have used the integral identity:
	\begin{align}
		\int_{0}^{2\pi}\frac{d\zeta}{2\pi}\frac{e^{im\zeta}}{A-\cos\zeta}=\frac{\left(A-\sqrt{A^2-1}\right)^{|m|}}{\sqrt{A^2-1}}.
	\end{align}
	The above identity is a deformation of the Gradshteyn and Ryzhik eight edition 3.613.

	The form factor part can be expressed by
	\begin{align}
		\mathcal{F}_{\vex{k},\vex{k}'}\mathcal{F}_{-\vex{k},-\vex{k}'}\approx \exp\left[-\mathcal{B}k_F^2+\mathcal{B}k_F^2e^{i\left(\theta_{\vex{k}}-\theta_{\vex{k}'}\right)}\right]=e^{-\mathcal{B}k_F^2}\left[\sum_{n=0}^{\infty}\frac{\left(\mathcal{B}k_F^2\right)^n}{n!}e^{in(\theta_{\vex{k}}-\theta_{\vex{k}'})}\right]\equiv \sum_{m}e^{im(\theta_{\vex{k}}-\theta_{\vex{k}'})}b_m(\mathcal{B}k_F^2),
	\end{align}
	where
	\begin{align}
		b_m(x\ge 0)=&\begin{cases}
			\frac{x^m}{m!}e^{-x} & \text{ for }m> 0,\\[1mm]
			e^{-x} & \text{ for }m= 0,\\[1mm]
			0 & \text{ for }m<0.
		\end{cases}
	\end{align}
	In this work, we focus only on $\mathcal{B}>0$, so $b_m(\mathcal{B}k_F^2)=0$ for all negative $m$. This is a special non-generic feature due to the ideal-quantum-geometry form factor considered in this work. Away from the ideal quantum geometry limit, $b_m(\mathcal{B}k_F^2)$ can be finite for all $m$, but $|b_{|m|}(\mathcal{B}k_F^2)|>|b_{-|m|}(\mathcal{B}k_F^2)|$ is generally true for $\mathcal{B}>0$.

	With the results above, we are in the position to derive $\mathcal{V}_L$, which is given by
	\begin{align}
		\nonumber\mathcal{V}(\nu_n;\vex{k},\vex{k}')\equiv& V(\nu_n,\vex{k}-\vex{k}')\mathcal{F}_{\vex{k},\vex{k}'}\mathcal{F}_{-\vex{k},-\vex{k}'}\approx g\sum_{m=-\infty}^{\infty}e^{im(\theta_{\vex{k}}-\theta_{\vex{k}'})}\left[\delta_{m,0}-R_m(\nu_n/\Omega_0)\right]
		\sum_{n=0}^{\infty}e^{in(\theta_{\vex{k}}-\theta_{\vex{k}'})}b_n(\mathcal{B}k_F^2)\\
		=&\sum_{L}e^{iL(\theta_{\vex{k}}-\theta_{\vex{k}'})}g\left[b_L(\mathcal{B}k_F^2)-\sum_nb_n(\mathcal{B}k_F^2)R_{L-n}(\nu_n/\Omega_0)\right]\equiv\sum_Le^{iL(\theta_{\vex{k}}-\theta_{\vex{k}'})}\mathcal{V}_L(\nu_n;k_F,\mathcal{B}).
	\end{align}
	where
	\begin{align}
		\mathcal{V}_L(\nu_n;k_F,\mathcal{B})=g\left[b_L(\mathcal{B}k_F^2)-\sum_nb_n(\mathcal{B}k_F^2)R_{L-n}(\nu_n/\Omega_0)\right].
	\end{align}

	For $\mathcal{B}=0$, the coefficient $b_n(0)=\delta_{n,0}$, resulting in $\mathcal{V}_{L}(\nu_n;k_F,\mathcal{B})=\mathcal{V}_{-L}(\nu_n;k_F,\mathcal{B})$. (Note that $R_L(x)=R_{-L}(x)$.)
	For $\mathcal{B}\neq 0$, we compare a few leading odd-$L$ pairing channels ($L=\pm 1,\pm 3$):
	\begin{align}
		\nonumber\mathcal{V}_1(\nu_n;k_F,\mathcal{B})=&g\left[b_1(\mathcal{B}k_F^2)-\sum_nb_n(\mathcal{B}k_F^2)R_{1-n}(\nu_n/\Omega_0)\right]\\
		=&g\left[b_1(\mathcal{B}k_F^2)-b_0(\mathcal{B}k_F^2)R_{1}(\nu_n/\Omega_0)-b_1(\mathcal{B}k_F^2)R_{0}(\nu_n/\Omega_0)-b_2(\mathcal{B}k_F^2)R_{-1}(\nu_n/\Omega_0)+\dots\right],\\
		\nonumber\mathcal{V}_{-1}(\nu_n;k_F,\mathcal{B})=&g\left[-\sum_nb_n(\mathcal{B}k_F^2)R_{-1-n}(\nu_n/\Omega_0)\right]\\
		=&g\left[-b_0(\mathcal{B}k_F^2)R_{-1}(\nu_n/\Omega_0)-b_1(\mathcal{B}k_F^2)R_{-2}(\nu_n/\Omega_0)-b_3(\mathcal{B}k_F^2)R_{-3}(\nu_n/\Omega_0)+\dots\right],\\
		\nonumber\mathcal{V}_3(\nu_n;k_F,\mathcal{B})=&g\left[b_3(\mathcal{B}k_F^2)-\sum_nb_n(\mathcal{B}k_F^2)R_{3-n}(\nu_n/\Omega_0)\right]\\
		=&g\left[b_3(\mathcal{B}k_F^2)-b_0(\mathcal{B}k_F^2)R_{3}(\nu_n/\Omega_0)-b_1(\mathcal{B}k_F^2)R_{2}(\nu_n/\Omega_0)-b_3(\mathcal{B}k_F^2)R_{1}(\nu_n/\Omega_0)+\dots\right],\\
		\nonumber\mathcal{V}_{-3}(\nu_n;k_F,\mathcal{B})=&g\left[-\sum_nb_n(\mathcal{B}k_F^2)R_{-3-n}(\nu_n/\Omega_0)\right]\\
		=&g\left[-b_0(\mathcal{B}k_F^2)R_{-3}(\nu_n/\Omega_0)-b_1(\mathcal{B}k_F^2)R_{-4}(\nu_n/\Omega_0)-b_3(\mathcal{B}k_F^2)R_{-5}(\nu_n/\Omega_0)+\dots\right].
	\end{align}
	For $L=1$ and $L=3$, the dominant contributions are due to the static attraction, i.e., $gb_L(\mathcal{B}k_F^2)$, which is completely absent for $L=-1$ and $L=-3$. The results straightforwardly indicate that $L>0$ pairings are dominant. The absence of static attraction for $L<0$ is due to the form factor $\mathcal{F}_{\vex{k},\vex{k}'}$ with ideal quantum geometry. Away from the ideal quantum geometry, finite static attraction can be induced in the $L<0$ channels, but the $L>0$ pairings are expected to be favorable.

	\section{Derivation of frequency-dependent linearized gap equation}

	Here, we discuss the derivation for the frequency-dependent linearized gap equation used in the main text. We consider imaginary-time action $\mathcal{S}_0+\mathcal{S}_I$, where
	\begin{align}
		\mathcal{S}_0=&\frac{1}{\beta}\sum_{\omega_n}\sum_{\vex{k}} \bar{c}_{\omega_n,\vex{k}}\left[-i\omega_n+E_{\vex{k}}-E_F\right]c_{\omega_n,\vex{k}},\\
		\mathcal{S}_{I}=&-\frac{1}{2\beta^3 \mathcal{A}}\sum_{\vex{k},\vex{k}'}\sum_{\omega_n,\omega_n'}\mathcal{V}(\omega_n-\omega_n';\vex{k},\vex{k}')\bar{c}_{\omega_n,\vex{k}}\bar{c}_{-\omega_n,-\vex{k}}c_{-\omega_n',-\vex{k}'}c_{\omega_n',\vex{k}'}\\
		\approx&-\frac{1}{2\beta^3 \mathcal{A}}\sum_{\vex{k},\vex{k}'}\sum_{\omega_n,\omega_n'}\sum_Le^{iL(\theta_{\vex{k}}-\theta_{\vex{k}'})}\mathcal{V}_L(\nu_n;k_F,\mathcal{B})\bar{c}_{\omega_n,\vex{k}}\bar{c}_{-\omega_n,-\vex{k}}c_{-\omega_n',-\vex{k}'}c_{\omega_n',\vex{k}'}\\
		\nonumber\rightarrow&-\frac{1}{2\beta}\sum_L\sum_{\omega_n}\sum_{\vex{k}}\left[\bar{\Delta}_L(\omega_n)e^{-iL\theta_{\vex{k}}}c_{-\omega_n,-\vex{k}}c_{\omega_n,\vex{k}}+e^{iL\theta_{\vex{k}}}\bar{c}_{\omega_n,\vex{k}}\bar{c}_{-\omega_n,-\vex{k}}\Delta_L(\omega_n)\right]\\
		&+\frac{\mathcal{A}\beta}{2}\sum_L\sum_{\omega_n,\omega_n'}\bar{\Delta}_L(\omega_n)\mathcal{V}_L^{-1}(\omega_n-\omega_n';k_F,\mathcal{B})\Delta_L(\omega_n'),
	\end{align}
	In the above expression, $\mathcal{V}_L^{-1}$ denotes the inverted matrix (with frequencies being the indexes) of $\mathcal{V}_L$, and 
	\begin{align} 
		\Delta_L(\omega_n)=&\frac{1}{\beta^2 \mathcal{A}}\sum_{\omega_n'}\sum_{\vex{k}'}\,\mathcal{V}_L(\omega_n\!-\!\omega_n';k_F,\mathcal{B})
		e^{-iL\theta_{\vex{k}'}}\!\left\langle c_{-\omega_n',-\vex{k}'}c_{\omega_n',\vex{k}'}\right\rangle.
	\end{align}

	Focusing the $L$ angular momentum pairing channel, the corresponding action is expressed by
	\begin{align}
		\nonumber\mathcal{S}_0+\mathcal{S}_{I}^{(L)}=&\frac{1}{\beta}\sum_{\omega_n}\sum_{\vex{k}}'\begin{bmatrix}
			\bar{c}_{\omega_n,\vex{k}}& c_{-\omega_n,-\vex{k}}
		\end{bmatrix}
		\begin{bmatrix}
			-i\omega_n+E_{\vex{k}}-E_F & -\Delta_L(\omega_n)e^{iL\theta_{\vex{k}}} \\
			-\bar{\Delta}_L(\omega_n)e^{-iL\theta_{\vex{k}}}& -i\omega_n-E_{-\vex{k}}+E_F
		\end{bmatrix}
		\begin{bmatrix}
			c_{\omega_n,\vex{k}}\\
			\bar{c}_{-\omega_n,-\vex{k}}
		\end{bmatrix}\\
		&+\frac{\mathcal{A}\beta}{2}\sum_{\omega_n,\omega_n'}\bar{\Delta}_L(\omega_n)\mathcal{V}_L^{-1}(\omega_n-\omega_n';k_F,\mathcal{B})\Delta_L(\omega_n'),
	\end{align}
	where the momentum space summation $\sum_{\vex{k}}'$ is over half of the available $k$ points (e.g., $k_x>0$) so that double counting is avoided.
	Next, we integrate out the fermionic fields in the partition function and derive the free energy density given by (with $\tilde{E}_{\vex{k}}=E_{\vex{k}}-E_F$ and infinitesimal $\Delta$)
	\begin{align}
		\mathcal{F}/\mathcal{A}=&-\frac{1}{\beta\mathcal{A}}\sum_{\omega_n}\sum_{\vex{k}}'\ln\left\{-\omega_n^2-\tilde{E}^2_{\vex{k}}-|\Delta_L|^2\right\}+\frac{1}{2}\sum_{\omega_n,\omega_n'}\bar{\Delta}_L(\omega_n)\mathcal{V}_L^{-1}(\omega_n-\omega_n';k_F,\mathcal{B})\Delta_L(\omega_n')\\
		\approx&-\frac{1}{\beta\mathcal{A}}\sum_{\omega_n}\sum_{\vex{k}}'\frac{|\Delta_L(\omega_n)|^2}{\omega_n^2+\tilde{E}^2_{\vex{k}}}+\frac{1}{2}\sum_{\omega_n,\omega_n'}\bar{\Delta}_L(\omega_n)\mathcal{V}_L^{-1}(\omega_n-\omega_n';k_F,\mathcal{B})\Delta_L(\omega_n')+\text{const}.
	\end{align}
	Requiring that the derivative of $\mathcal{F}/\mathcal{A}$ with respect to $\bar\Delta_L(\omega_n)$ vanishes, we obtain the frequency-dependent linearized gap equation is given by
	\begin{align}
		\Delta_L(\omega_n)=\frac{2}{\beta \mathcal{A}}\sum_{\omega_n'}\sum_{\vex{k}'}'\frac{\mathcal{V}_L(\omega_n-\omega_n';k_F,\mathcal{B})}{(\omega_n')^2+\tilde{E}^2(\vex{k}')}\Delta_L(\omega_n')=\frac{1}{\beta}\sum_{\omega_n'}\int_{-\Lambda'}^{\Lambda'} dE \frac{\rho(E)}{(\omega_n')^2+E^2}\mathcal{V}_L(\omega_n-\omega_n';k_F,\mathcal{B})\Delta_L(\omega_n'),
	\end{align}
	where $\Lambda'$ is the energy cutoff. 
	For a system with a constant density of states, we replace $\rho(E)$ by $\rho(\mu)\equiv \rho_0$. In the limit $\Lambda'\gg\Omega_0$, we obtain the following asymptotic expression
	\begin{align}
		&\frac{\rho_0}{\beta}\sum_{\omega_n'}\int_{-\Lambda'}^{\Lambda'} dE \frac{1}{(\omega_n')^2+E^2}\mathcal{V}_L(\omega_n-\omega_n';k_F)\Delta_L(\omega_n')\bigg|_{\Lambda'\rightarrow\infty}=\frac{\rho_0\pi}{\beta}\sum_{\omega_n'}\frac{\mathcal{V}_L(\omega_n-\omega_n';k_F,\mathcal{B})}{|\omega_n'|}\Delta_L(\omega_n')\\
		\rightarrow&\Delta_L(\omega_n)\approx\frac{\rho_0\pi}{\beta}\sum_{\omega_n'}\frac{\mathcal{V}_L(\omega_n-\omega_n';k_F)}{|\omega_n'|}\Delta_L(\omega_n')\equiv\lambda\sum_{\omega_n'}\frac{\pi\left[\mathcal{V}_L(\omega_n-\omega_n';k_F,\mathcal{B})/g\right]}{\beta|\omega_n'|}\Delta_L(\omega_n'),
	\end{align}
	where $\lambda=g\rho_0$ is the dimensionless coupling constant. 
	The last equation recasts the linearized gap equation as an eigenvalue problem with matrix indices corresponding to the Matsubara frequencies. The temperature ($T=\beta^{-1}$) controls the spacing in the Matsubara frequencies. The critical temperature $T_c$ corresponds to $T$ such that the right eigenvalue is 1. (Note that the matrix is not symmetric, so we focus only on the right eigenvalue.)
	
	%In practice, we need to set a frequency cutoff $\Lambda$ (not the energy cutoff $\Lambda'$) in the numerical calculations. We choose $\Lambda\ge 5\Omega_0$ as the results converge well for $\Lambda>5\Omega_0$. The convergence behavior is not surprising given the rapid falloff of the pairing interaction at high frequencies. It is more convenient to fix a $T$ and determine the corresponding $\lambda_c$. For a fixed $\lambda$ calculation, we need to dynamically tune the number of kept frequencies such that $\Lambda\ge 5\Omega_0$ is valid. We discuss the convergence in depth in the next section.
	
	\section{Numerical solutions of linearized gap equation}
	
	\begin{table*}[t]

		%----------------------------------%
		% Left table
		%----------------------------------%
		\begin{minipage}[t]{0.3\textwidth}
			\centering
			
			\textbf{$T = 0.001\Omega_0$}
			
			\vspace{0.3em}
			\begin{tabular}{c c c }
				\hline\hline
				$N$ &   $\Lambda/\Omega_0$ & $\lambda(N)$ \\
				\hline
				800  &  2.51   &  3.25304 \\
				1600 &  5.02  & 3.24390 \\
				2400 &  7.54  & 3.24324 \\
				3200 &  10.05  & 3.24310 \\
				4000 &  12.56  & 3.24306	\\
				\hline\hline
			\end{tabular}
		\end{minipage}
		\hfill
		%----------------------------------%
		% Center table
		%----------------------------------%
		\begin{minipage}[t]{0.3\textwidth}
			\centering
			
			\textbf{$T = 0.0005\Omega_0$}
			
			\vspace{0.3em}
			\begin{tabular}{c c c}
				\hline\hline
				$N$ & $\Lambda/\Omega_0$ & $\lambda(N)$  \\
				\hline
				2000  & 3.14 &  2.93708 \\
				4000 & 6.28 & 2.93299  \\
				6000 & 9.42 & 2.93269 \\
				8000 & 12.56  & 2.93263 \\
				\hline\hline
			\end{tabular}
		\end{minipage}
		\hfill
		%----------------------------------%
		% Right table
		%----------------------------------%
		\begin{minipage}[t]{0.3\textwidth}
			\centering
			
			\textbf{$T = 0.0001\Omega_0$}
			
			\vspace{0.3em}
			\begin{tabular}{c c c}
				\hline\hline
				$N$ & $\Lambda/\Omega_0$ & $\lambda(N)$  \\
				\hline
				8000  & 2.51 & 2.43802 \\
				16000 & 5.03 & 2.42887 \\
				24000 & 7.54 & 2.42824 \\
				32000 & 10.05 & 2.42811 \\
				\hline\hline
			\end{tabular}
		\end{minipage}
		\caption{Finite-size dependence of dimensionless coupling constant with three different temperatures. Left: $T=0.001\Omega_0$. Center: $T=0.0005\Omega_0$. Right: $T=0.0001\Omega_0$. $N$ denotes the number of Matsubara frequencies; $\Lambda=2\pi T\frac{(N-1)}{2}$ is the frequency cutoff; $\lambda(N)$ is the extracted dimensionless coupling constant from solving the linearized gap equation for $L=1$ and $\mathcal{B}=0$ with $N$ frequencies.}
		\label{tab:eigenvalues_vs_N}
		
	\end{table*}
	
	Here, we discuss the convergence issue in solving the linearized gap equation. In the numerical calculations, we need to assign a finite number of Matsubara frequencies ($N$), corresponding to a frequency cutoff, $\Lambda=2\pi T\frac{(N-1)}{2}$ (not the energy cutoff $\Lambda'$), where we always consider an even number $N$. We test the results as functions of $N$ below.
	
	First, we rewrite the linearized gap equation as an explicit eigenvalue problem as follows:
	\begin{align}\label{SEq:LGE_eigen}
		\sum_{\omega_n'}M_{\omega_n,\omega_n'}\Delta_L(\omega_n')=\frac{1}{\lambda}\Delta_L(\omega_n).
	\end{align}
	where 
	\begin{align}
		M_{\omega_n,\omega_n'}=\pi T\frac{\left[\mathcal{V}_L(\omega_n-\omega_n';k_F,\mathcal{B})/g\right]}{|\omega_n'|}.
	\end{align}
	Since $M_{\omega_n,\omega_n'}$ is not symmetric, $1/\lambda$ and $\Delta_L(\omega_n)$ correspond to the right eigenvalue and the right eigenvector, respectively. For a given $T$ and $N$, one can obtain $\lambda$ and $\Delta_L(\omega_n)$ through numerical diagonalization. Thus, $\lambda$ and $\Delta_L(\omega_n)$ depend on both $T$ and $N$. We investigate the convergence of both quantities $L=1$ and $\mathcal{B}=0$.
	
	We investigate the convergence of $\lambda$. In Table~\ref{tab:eigenvalues_vs_N}, we list $\lambda(N)$ with three different temperatures, $T=0.001\Omega_0$, $T=0.0005\Omega_0$, and $T=0.0001\Omega_0$ (the lowest temperature used in this work). We find that the value of $\lambda$ converges well (relative error $<10^{-4}$) for a sufficiently large $N$ such that $\Lambda>5\Omega_0$. Interestingly, for $\Lambda\approx 2.5\Omega_0$, the extracted $\lambda$ value is still very close to the convergent value (within 0.5 percent). The convergence behavior is not surprising given the rapid falloff of the pairing interaction at frequencies higher than $5\Omega_0$.
	
\begin{figure}[t!]
		\includegraphics[width=0.4\textwidth]{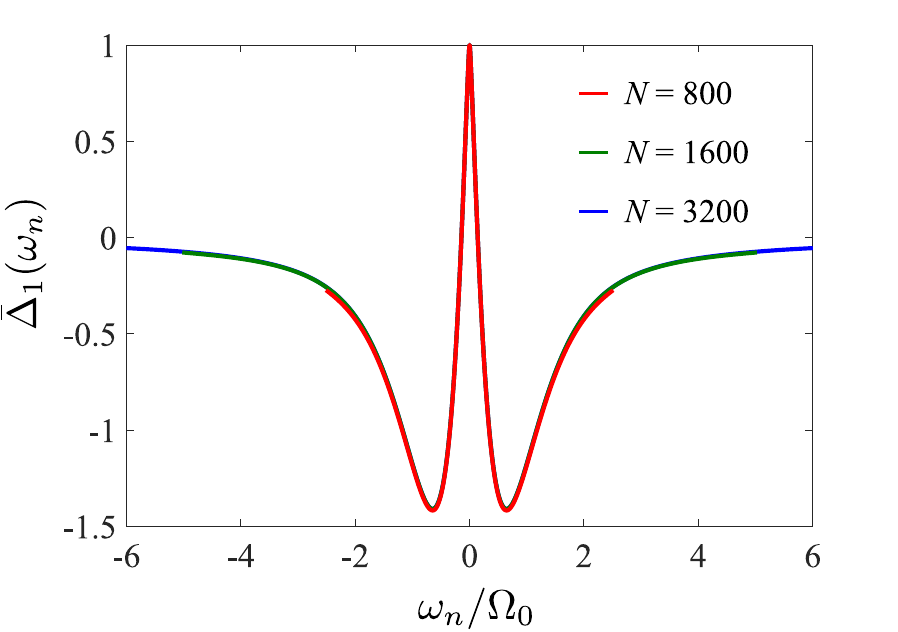}
		\caption{The gap function with different numbers of Matsubara frequencies. $\bar{\Delta}_1(\omega_n)\equiv\Delta_L(\omega_n)/\Delta_1(\pi T)$ is the rescaled gap function for $L=1$ and $\mathcal{B}=0$. We set $T=10^{-3}\Omega_0$ and vary the number of Matsubara frequencies $N$.
		}
		\label{Fig:finite_N_Delta}
	\end{figure}

	Next, we investigate the convergence of the gap function. In Fig.~\ref{Fig:finite_N_Delta}, we plot the rescaled gap function for $L=1$, $\mathcal{B}=0$, and $T=10^{-3}\Omega_0$ with $N=800$ ($\Lambda\approx 2.5\Omega_0$), $N=1600$ ($\Lambda\approx 5.02\Omega_0$), and $N=3200$ ($\Lambda\approx 10.05\Omega_0$). The $N=800$ case shows small deviations from other cases, and the curves of $N=1600$ and $N=3200$ are almost identical. Similar convergence behavior is found for other cases, indicating that the gap function also converges well for $\Lambda>5\Omega_0$.

	In conclusion, the numerical results here show that both $\lambda$ and $\Delta_L$ converge well for a sufficiently large $N$ such that $\Lambda>5\Omega_0$. So far, we consider only the case with $\mathcal{B}=0$ and $L=1$. The same is true for $\mathcal{B}\neq 0$ and/or $L\neq 1$. The convergence can be understood by the pairing interactions in Figs. 2(c) (for $L=3$), 2(d) (for $L=5$), and 3(a) (for $\mathcal{B}\neq 0$), where the tails of the pairing interactions (e.g., $|\omega_n|>2\Omega_0$) are even smaller than the case with $L=1$ and $\mathcal{B}=0$. In fact, these cases converge better than case with $L=1$ and $\mathcal{B}=0$.
	All the numerical results in this work are obtained under the condition $\Lambda>5\Omega_0$.

	\section{$\mathcal{B}k_F^2$ in rhombohedral graphene multilayers}

	In this section, we discuss the quantity $\mathcal{B}k_F^2$ in the realistic rhombohedral graphene multilayers, specifically four to six layers. As discussed in the main text, $\mathcal{B}k_F^2$ is essential for the phonon-mediated SC as it can enhance the superconducting $T_c$ and induce higher-angular-momentum pairings. We focus on the experimental conditions of the observed quarter-metal superconductivity \cite{HanT2025a,MorissetteE2025b}. Note that the hexalayer experiment suggests an unusual striped superconductivity \cite{MorissetteE2025b}, rather than the chiral superconductivity observed for the tetralayer and pentalayer \cite{HanT2025a}. However, the shape of the two superconducting regions in the hexalayer and the corresponding normal-state properties are qualitatively consistent with the tetralayer and the pentalayer experiments. We thus include the hexalayer case in the discussion. Future experiments should clarify the precise nature and the phase diagram in the rhombohedral hexalayer graphene.
	
\begin{figure}[t!]
		\includegraphics[width=1\textwidth]{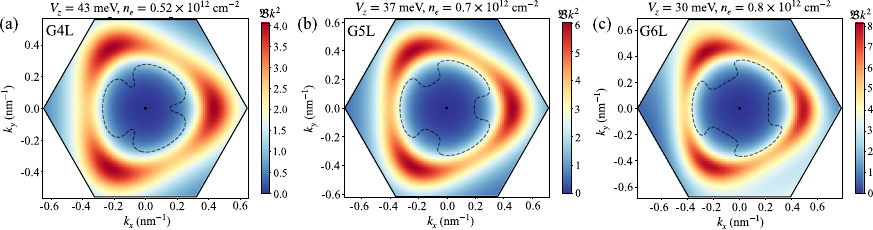}
		\caption{$\mathcal{B}k_F^2$ for rhombohedral graphene multilayers using $k\cdot p$ model. We compute $\mathcal{B}k^2$ (with $k$ relative to the valley momentum) and draw Fermi surface contours (focusing on a single Fermi surface) for the doping density within experimentally superconducting regions. (a) Tetralayer with $V_z=43$ meV and $n_e=0.52\times10^{12}$ cm$^{-2}$. (b) Pentalayer with $V_z=37$ meV and $n_e=0.7\times10^{12}$ cm$^{-2}$. (a) Hexalayer with $V_z=30$ meV and $n_e=0.8\times10^{12}$ cm$^{-2}$. We also average $\mathcal{B}k_F^2$ over the $k$ points near the Fermi surfaces (denoted by $\overline{\mathcal{B}k_F^2}$) for each cases and obtain $\overline{\mathcal{B}k_F^2}\approx 0.47$ for $n=4$ (a), $\overline{\mathcal{B}k_F^2}\approx 1.14$ for $n=5$ (b), and $\overline{\mathcal{B}k_F^2}\approx 1.5$ for $n=6$ (c). Note the different color bar scales in each figure.
		}
		\label{Fig:Bk2}
	\end{figure}

	We employ the $k\cdot p$ model with the band parameters in Ref.~\cite{GhazaryanA2023}. In Fig.~\ref{Fig:Bk2}, we plot $\mathcal{B}k^2$ for tetralayer ($n=4$), pentalayer ($n=5$), and hexalayer ($n=6$), with the value of $V_z$ (the electric potential difference between the adjacent layers) consistent with the experiments \cite{HanT2025a,MorissetteE2025b}. Notably, the realistic rhombohedral graphene multilayers do not have ideal circular Fermi surfaces, and the Fermi surfaces associated with the Van Hove singularity are rather complex. Thus, we focus only on the single-Fermi-surface cases with approximately circular Fermi surfaces [the black-dashed lines in Fig.~\ref{Fig:Bk2}].  
	The value of $\mathcal{B}k^2$ is of order one for most of the $k$ points near the Fermi surface. We also average $\mathcal{B}k_F^2$ over the $k$ points near the Fermi surfaces (denoted by $\overline{\mathcal{B}k_F^2}$) for each cases and obtain $\overline{\mathcal{B}k_F^2}\approx 0.47$ for $n=4$, $\overline{\mathcal{B}k_F^2}\approx 1.14$ for $n=5$, and $\overline{\mathcal{B}k_F^2}\approx 1.5$ for $n=6$. The results indicate a monotonically increasing $\overline{\mathcal{B}k_F^2}$ as $n$ increases. Within the parameter regime of interest, the Berry curvature is insensitive to the value of $V_z$. 
	
	Next, we discuss the implications of our theory to the superconductivity in the existing rhombohedral graphene multilayers. First, we estimate the dimensionless coupling constant $\lambda$ for $n=4$. Using $g=0.474$ eV nm$^2$ \cite{WuF2019,ChouYZ2021} and $\rho_0=6.6$ eV$^{-1}$ nm$^{-2}$ [corresponding to the Fermi surface in Fig.~\ref{Fig:Bk2}(a)], we obtain $\lambda=g\rho_0\approx3.13$, an order-one coupling constant. The value $\lambda\approx3.13$ here is not precise, as a significantly larger $\lambda$ can be obtained by selecting a slightly smaller $n_e$ (toward the Van Hove singularity). With the estimated $\lambda$ and the averaged $\mathcal{B}k_F^2$ (i.e., $\overline{\mathcal{B}k_F^2}$ discussed in the previous paragraph), we expect that the phonon-mediated pairing is likely important for superconductivity in the rhombohedral tetralayer graphene. Crucially, the maximal $T_c$ should roughly track the maximal $\rho(E_F)\mathcal{V}_L(0;k_F,\mathcal{B})$ (not the $\lambda=g\rho_0$ introduced in the main text and discussed previously).

	We caution that vertex corrections may be important for rhombohedral graphene systems, as the Fermi velocity is comparable to or smaller than the sound velocity \cite{ChouYZ2022b}, potentially invalidating the Migdal theorem. Another outstanding issue is that the realistic Fermi surface is not circular. Thus, estimating $T_c$ directly using our theory is not quantitatively reliable, and the future microscopic calculations should incorporate the $k\cdot p$ model dispersion as well.
	
	Regardless of the quantitative issue about the precise $T_c$, we point out two qualitative predictions based on our theory and the numerical results presented here. First, the value of $\overline{\mathcal{B}k_F^2}$ increases monotonically as the number of layers increases, suggesting possible $L>1$ pairings for $n>7$. In principle, $L>1$ pairings can happen in $n=4$ already, provided that the doping density is high enough to reach the ring of intensive Berry curvature \cite{PatriAS2025b}. However, the corresponding $\lambda$ is likely parametrically small, i.e., unable to support observable superconductivity. Second, the pairing interactions in our theory are proportional to $\exp(-\mathcal{B}k_F^2)$ upto some power-law corrections, indicating that the pairing is strongly suppressed for a large $\mathcal{B}k_F^2$, which is likely the case for systems with a high layer number $n$. Therefore, the quarter-metal superconductivity is likely observable only for a range of layers with $n\ge 4$, not all the way to the rhombohedral graphene limit. 
	
\end{document}